\documentclass[twocolumn,showpacs,preprintnumbers,amsmath,amssymb]{revtex4-1}

\usepackage{graphicx}
\usepackage{dcolumn}
\usepackage{bm}
\usepackage{epstopdf}
\usepackage{amsmath,amsbsy,amsfonts,amssymb}
\usepackage{epsfig}

\begin{document}
\title{Energy and radiative properties of the (3)$^1\Pi$ and (5)$^1\Sigma^+$ states of RbCs. Experiment and theory.}
\author{K. Alps, A. Kruzins, O, Nikolayeva,  M. Tamanis, and R. Ferber}
\affiliation{Laser Center, University of Latvia, 19 Rainis Boulevard, Riga LV-1586, Latvia}
\email{ferber@latnet.lv}
\author{E. A. Pazyuk and  A. V. Stolyarov}
\affiliation{Department of Chemistry, Lomonosov Moscow State University, 119991 Moscow, Leninskie gory 1/3, Russia}
\email{avstol@phys.chem.msu.ru}
\date{\today }

\begin{abstract}
We combined high resolution Fourier-transform spectroscopy and large scale electronic structure calculation to study energy and radiative properties of the high-lying (3)$^1\Pi$ and (5)$^1\Sigma^+$ states of the RbCs molecule. The laser induced (5)$^1\Sigma^+$,(4)$^1\Sigma^+$,(3)$^1\Pi\to$A(2)$^1\Sigma^+\sim $~b(1)$^3\Pi$ fluorescence (LIF) spectra were recorded by the Bruker IFS-125(HR) spectrometer in the frequency range $\nu\in [5500,10000]$ cm$^{-1}$ with the instrumental resolution of 0.03 cm$^{-1}$. The rotational assignment of the observed LIF progressions, which exhibit irregular vibrational - rotational spacing due to strong spin-orbit interaction between A$^1\Sigma^+$ and b$^3\Pi$ states was based on the coincidences between observed and calculated energy differences. The required rovibronic term values of the strongly perturbed A$\sim$b complex have been calculated by coupled - channel approach for both $^{85}$Rb$^{133}$Cs and $^{87}$Rb$^{133}$Cs isotopolgues with accuracy of about 0.01 cm$^{-1}$, as demonstrated in [A. Kruzins et. al., J. Chem. Phys. 141,184309 (2014)]. The experimental energies of the upper (3)$^1\Pi$ and (5)$^1\Sigma^+$ states were involved in a direct-potential-fit analysis performed in the framework of inverted perturbation approach. Quasi-relativistic \emph{ab initio} calculations of the spin-allowed (3)$^1\Pi$,(5)$^1\Sigma^+\to$ (1-4)$^1\Sigma^+$,(1-3)$^1\Pi$ transition dipole moments were performed. Radiative lifetimes and vibronic branching ratios of radiative transitions from (3)$^1\Pi$ and (5)$^1\Sigma^+$ states were evaluated. To elucidate the origin of the $\Lambda$-doubling effect in the (3)$^1\Pi$ state the angular coupling (3)$^1\Pi$-(1-5)$^1\Sigma^+$ electronic matrix elements were calculated and applied for the relevant $q-$factors estimate. The intensity distributions simulated for the particular (5)$^1\Sigma^+$;(3)$^1\Pi\to$A$\sim$b LIF progressions have been found to be remarkably close to their experimental counterparts.
\end{abstract}
\maketitle

\section{Introduction}

Accurate knowledge of energy and radiative properties of a high lying excited state of heavy polar alkali diatomic molecules is important for selecting the efficient optical routes for accessing so-called intermediate electronic states of the mixed singlet-triplet nature in one- or two-photon transitions~\cite{Bergmann2015, RCR_Obzor15}; besides, it provides an unambiguous test of the reliability of state-of-the-art \emph{ab initio} electronic structure calculations~\cite{Tennyson2016}.

\begin{figure}
\includegraphics[scale=0.9]{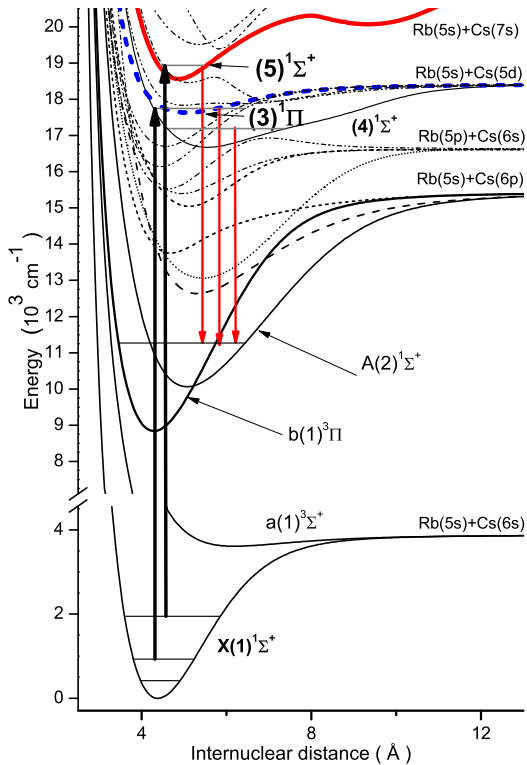}
\caption{Scheme of the lowest electronic terms of  RbCs molecule~\cite{Allouche00}. Vertical arrows mark excitation - observation transitions in the  present FTS LIF experiment.}\label{Fig1_PECs}
\end{figure}

In particular, the RbCs molecule has been actively studied both theoretically and experimentally because of its successful employment for obtaining ultracold polar molecular species Refs~\cite{Takekoshi2014, Molony2014, DeMille2014, DeMille2016, Qu_ChRev2012}. At the same time the accurate empirical information on the higher electronic states of RbCs is still very scarce. To our best knowledge, the only electronic state approaching the atomic limit higher than 5$^2$P(Rb)+6$^2$S(Cs), which was studied experimentally with high resolution in broad enough range of vibrational $v^{\prime}$ and rotational $J^{\prime}$ levels, is the (4)$^1\Sigma^+$ state~\cite{Nogueira2008, ZutersPRA2013} (see, Fig.~\ref{Fig1_PECs}); this state appeared to be promising for two-step optical cycles to achieve ultracold RbCs in the absolute ($v_X=0$, $J_X=0$) ground state because of efficient transitions to the lowest electronic X$^1\Sigma^+$ and a$^3\Sigma^+$ states.

The information on other high-lying states of RbCs is fragmentary, since the main goal often was to use the subsequent laser induced fluorescence (LIF) to study the ground X$^1\Sigma^+$ state. For instance, molecular constants (Dunham coefficients) and corresponding Rydberg-Klein-Rees (RKR) potentials for (2,4,5)$^1\Pi$ and (3,7)$^1\Sigma^+$ states of RbCs have been obtained~\cite{Gustavsson1988} by Fourier Transform Spectroscopy (FTS) method in the energy regions 20000 - 22000 cm$^{-1}$ and 13000 - 15000 cm$^{-1}$. As far as high resolution FTS of LIF from high - lying electronic states to the ground  state is concerned, the main difficulties are often connected with the low probability of optical transitions due to a small transition electric dipole moment. In many cases even if a low probability transition may still be used to excite a high lying state, it is hopeless to convincingly detect dispersed LIF backwards to the ground X$^1\Sigma^+$  state. Therefore, the alternative methods might be preferable, such as  polarization labeling spectroscopy, two-photon optical transitions, or multi-photon ionization. Information on the (4)$^1\Sigma^+$, (3)$^1\Pi$, (5)$^1\Sigma^+$, (6)$^1\Sigma^+$ and (4)$^3\Pi$ states obtained from the high-resolution resonance-enhanced two-photon ionization (RE2PI) spectroscopy of RbCs in a molecular beam is contained in Refs.\cite{Kim93, Kim94, Yoon-JCP2001, Kim2009}; the spectroscopic data on (5)$^1\Sigma^+$ and (3)$^3\Pi$ states can be found in Ref.~\cite{Yoon-JCP2001}.

In the present study we suggest an alternative method to overcome weak LIF to the ground state: to replace the latter by the first excited state, namely by the LIF transitions to the strongly mixed A(2)$^1\Sigma^+\sim $b(1)$^3\Pi$ system (see, Fig.~\ref{Fig1_PECs}). This opportunity has been made feasible by the recent high accuracy deperturbation analysis of the RbCs $A\sim b$ complex~\cite{KruzinsRbCs2014}, which allows one to reproduce the required rovibronic term values of the complex with the accuracy quite comparable with that of the ground state. We focused our study on the (3)$^1\Pi$ and (5)$^1\Sigma^+$ states of RbCs with the purpose to obtain more detailed energy information needed to perform the direct-potential-fit (DPF) analysis of these states. We also performed \emph{ab initio} calculations of potential energy curves (PECs) and transition  dipole moments to estimate radiative properties of both states under study.

\begin{figure}
\includegraphics[scale=0.9]{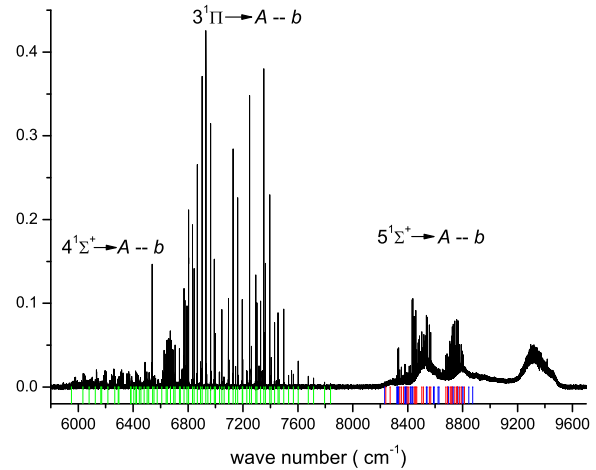}
\caption{Example of FTS LIF spectrum to the A$\sim$b complex recorded at excitation laser frequency 17672.697 cm$^{-1}$ using long-pass edge filter 1050 nm. Green bars mark the singlet $Q$-transitions from the (3)$^1\Pi$ level corresponding to $v^{\prime} = 14$, $J^{\prime} = 73$ and $E^{\prime}=$ 17787.008 cm$^{-1}$; red and blue bars mark the doublet $P/R$-transitions from the (5)$^1\Sigma^+$ levels with $v^{\prime} = 5$, $J^{\prime}= 211$, $E^{\prime}=$ 19346.149 cm$^{-1}$ and $v^{\prime} = 7$, $J^{\prime} = 97$, $E^{\prime}=$ 18974.387 cm$^{-1}$, respectively.}\label{Fig2_spectra}
\end{figure}

\section{Experiment}\label{experiment}

In the experiment RbCs molecules were produced in the heat-pipe oven, which was loaded with 10g Cs and 5g Rb (natural mixture of isotopes) metals. Working temperature was 310 $^\circ$C. For excitation the single mode ring dye laser Coherent 699-21 with Rhodamine 6G dye was used. Laser beam was sent into the heat-pipe and backward LIF spectra to the strongly mixed A$^1\Sigma^+$ and b$^3\Pi$ states from the upper (3)$^1\Pi$ and (5)$^1\Sigma^+$ states, as well as from accidentally excited (4)$^1\Sigma^+$ state, see Fig.~\ref{Fig1_PECs}, were recorded with the instrumental resolution of 0.03 cm$^{-1}$ by means of FT spectrometer Bruker IFS-125(HR). LIF was detected in the range from 5500 to 10000 cm$^{-1}$ using InGaAs detector. Laser frequency was selected within the range 16700 - 17800 cm$^{-1}$. This frequency range is very suitable to excite (4)$^1\Sigma^+$ state, which typically gives rise to a strong (4)$^1\Sigma^+\to$A$\sim$b fluorescence ~\cite{KruzinsRbCs2014}. Fortunately, the (3)$^1\Pi\to$A$\sim$b system is partially shifted from (4)$^1\Sigma^+\to$A$\sim$b system due to a different equilibrium internuclear distance of the upper state PECs. Thus, by monitoring in the expected range the (3)$^1\Pi\to$A$\sim$b LIF signal in real time by means of FT spectrometer's Preview Mode at low resolution, it was possible to set the laser frequency in resonance with the (3)$^1\Pi(v^{\prime},J^{\prime}) \leftarrow$X$^1\Sigma^+(v^{\prime\prime},J^{\prime\prime})$ transitions. An example of the recorded spectrum is shown in Fig.~\ref{Fig2_spectra}. In this particular case an accidentally excited (4)$^1\Sigma^+\to$A$\sim$b system is strongly diminished, when compared with the strongest part of the (3)$^1\Pi\to$A$\sim$b system in the range 6800 - 7600 cm$^{-1}$. The LIF  (5)$^1\Sigma^+\to$A$\sim$b is observed in the range 8400 - 8800 cm$^{-1}$ and is typically weaker than the (3)$^1\Pi\to$A$\sim$b system, at least for two reasons. First, as one can see in Fig.~\ref{Fig1_PECs} the (5)$^1\Sigma^+$ state lies about 1500 cm$^{-1}$ higher than the (3)$^1\Pi$ state; hence, the absorption transitions start from high, less populated vibrational $v^{\prime\prime}$ levels of the ground state. Another reason is the continuum fluorescence observed in the range 8400 - 9600 cm$^{-1}$, which is overlapping with the (5)$^1\Sigma^+\to$A$\sim$b LIF. Since in the Preview Mode the continuum LIF is dominating, it was practically impossible to set the laser frequency in resonance to excite the (5)$^1\Sigma^+$ state. Nevertheless, we have managed to record more than 100 LIF progressions from (3)$^1\Pi$ and also (5)$^1\Sigma^+$ states to the A$\sim$b complex, which exhibited an irregular wibrational - rotational spacing.

\section{Rotational assignment of the of irregular LIF progressions to the A$\sim$b complex}

Assignment of the observed (5)$^1\Sigma^+$, (3)$^1\Pi\to$ A$^1\Sigma^+\sim$~b$^3\Pi$ transitions was based on the high accuracy description of the A$\sim$b complex ~\cite{KruzinsRbCs2014}, which ensures that its rovibronic term values can be reproduced with accuracy typically better than 0.01 cm$^{-1}$ in wide  energy range.

The required rovibronic energy $E_{A\sim b}$ and relevant non-adiabatic vibrational wave function $\mathbf{\Phi}_{A\sim b}$ of the A$\sim$b complex were generated in advance for both $^{85}$Rb$^{133}$Cs and $^{87}$Rb$^{133}$Cs isotopologues by the numerical solution of the coupled-channel (CC) radial equations~\cite{Bergeman2003}:
\begin{eqnarray}\label{CC}
\left(- {\bf I}\frac{\hbar^2 d^2}{2\mu dR^2} + {\bf V}(R;\mu,J^{\prime\prime}) -
{\bf I}E_{A\sim b}\right)\mathbf{\Phi_{A\sim b}}(R) = 0
\end{eqnarray}
where ${\bf I}$ is the identity matrix, $\mu$ is the reduced mass and the $4\times 4$ symmetric matrix ($i\in [A^1\Sigma^+, b^3\Pi_{\Omega = 0,1,2}]$) of the potential energy ${\bf V}(R;\mu,J^{\prime\prime})$ consists of nonzero diagonal
\begin{eqnarray}\label{diagonalAb4}
V_{^1\Sigma^+} &=& U_A+B[X+2]\\
V_{^3\Pi_0} &=& U_{b0}+B[X+2]\nonumber\\
V_{^3\Pi_1} &=& U_{b0}+A_{01}+B[X+2]\nonumber\\
V_{^3\Pi_2} &=& U_{b0}+2A_{so}+B[X-2]\nonumber
\end{eqnarray}
and off-diagonal
\begin{eqnarray}\label{offdiagonalAb4}
V_{^1\Sigma^+-^3\Pi_0}&=&-\sqrt{2}\xi_{Ab0}\\
V_{^3\Pi_0-^3\Pi_1} &=&-B\sqrt{2X}\nonumber\\
V_{^3\Pi_1-^3\Pi_2} &=&-B\sqrt{2(X-2)}\nonumber
\end{eqnarray}
matrix elements with
\begin{eqnarray}\label{B-rotational}
B=\frac{\hbar^2}{2\mu R^2}&;& \quad X\equiv J^{\prime\prime}(J^{\prime\prime}+1).
\end{eqnarray}
The rotationless $U_A(R)$ and $U_{b0}(R)$ PECs of the mutually perturbed singlet and triplet states as well as the corresponding spin-orbit coupling $\xi_{Ab0}(R)$ and the non-equidistant SO splitting $A_{01}(R)\neq A_{so}(R)$ functions were borrowed from Ref.\cite{KruzinsRbCs2014}. The resulting term values data set of the A$\sim$b complex is provided in the Supplemented material~\cite{EPAPS}.

The rotational assignment of  the lines belonging to a particular LIF progression was performed in automatic regime with help of a homemade programme, which searched for the coincidences between calculated and observed rovibrational differences. Two criteria served as a test of the correctness of the assignment. First, the assigned transitions belonging to a particular progression should yield to a coinciding, within a given accuracy, upper state energy. Second, the excitation laser frequency should be in resonance, within Doppler broadening, with some calculated absorption transition from the ground state level. Moreover, all (4)$^1\Sigma^+\rightarrow$A$\sim$b transitions observed in the recorded LIF spectra were assigned too, and the obtained (4)$^1\Sigma^+$ state term values were compared with the ones calculated by the empirical PEC from Ref.~\cite{ZutersPRA2013}. These tests unambiguously proved the reliability of the assigned procedure.

\section{Direct-potential-fit (DPF) analysis}\label{DPF}

The present experimental term values of the (3)$^1\Pi$ and (5)$^1\Sigma^+$ states were incorporated, together with the preceding experimental data~\cite{Kim94, Yoon-JCP2001}, in the DPF analysis performed in the framework of inverted perturbation approach~\cite{IPA1} (IPA). The adjusted adiabatic potentials $U^{IPA}_i(R)$ of both studied states were defined in the point-wise interpolated by the natural cubic splines on a finite interval of internuclear distance $R$. The required initial points were extracted from the corresponding "difference-based" potentials $U^{db}_i(R)$ constructed by Eq.(\ref{difbased}) as described in Section~\ref{abinitio}.

In order to take into account the $\Lambda$-doubling effect in the (3)$^1\Pi$ state explicitly, the conventional effective (rotational) potential used in the iteration solution of the radial Schr\"{o}dinger equation was modified as
\begin{eqnarray}\label{centrifugal}
U^{J^{\prime}} = U^{IPA} + (1 + sBQ)B[J^{\prime}(J^{\prime}+1)-1],
\end{eqnarray}
where $B(R)$ is given by Eq.(\ref{B-rotational}) while $Q(R)$ is the \emph{ab initio} precomputed function of the internuclear distance:
\begin{eqnarray}\label{Qfunction}
Q(R)=2\sum_{^1\Sigma^+}\frac{|L^{ab}_{^1\Pi-^1\Sigma^+}|^2}{U^{ab}_{^1\Pi}-U^{ab}_{^1\Sigma^+}}.
\end{eqnarray}
The required potentials $U_i^{ab}(R)$ and $L$-uncoupling electronic matrix elements $L^{ab}_{ij}(R)$ between the (3)$^1\Pi$ and (1-5)$^1\Sigma^+$ states were obtained during the electronic structure calculation described in Section~\ref{abinitio}. The scaling parameter $s$ in Eq.(\ref{centrifugal}) apparently vanishes for the $f$-component of the (3)$^1\Pi$ state whereas for the $e$-component $s$ is considered as the adjustable fitting parameter, which should be close to one.

\section{Electronic structure calculation}\label{abinitio}

The adiabatic PECs $U^{ab}_i(R)$ for the $i\in $(1-5)$^1\Sigma^+$, (1-4)$^1\Pi$ states, along with a spin-allowed $^1\Sigma^+-^1\Sigma^+$ and $^1\Pi-^1\Sigma^+$ transition dipole moments $d_{ij}^{ab}(R)$, were calculated in the basis of spin-averaged electronic wavefunctions corresponding to pure (\textbf{a}) Hund's coupling case. To elucidate an origin of the $\Lambda$-doubling effect observed in the (3)$^1\Pi$ state the relevant angular coupling electronic matrix elements $L_{ij}^{ab}(R)$ between the (3)$^1\Pi$ state and (1-5)$^1\Sigma^+$ states were evaluated as well. All calculations were performed in the wide range of internuclear distance $R\in [2.5,25]$~\AA~by means of the MOLPRO program~\cite{MOLPRO}.

Assuming that the originally calculated electronic energy weakly depends on the excitation energy, the $R$-dependent part of a systematic error in the original \emph{ab initio} curves $U_i^{ab}(R)$ was decreased for the excited states by means of a semi-empirical expression~\cite{Zaitsevskii05}:
\begin{eqnarray}\label{difbased}
U_i^{db} = [U_i^{ab}-U_X^{ab}] + U_X^{emp},
\end{eqnarray}
where $U_X^{emp}(R)$ is the highly accurate empirical PEC available for the ground X$^1\Sigma^+$ state~\cite{Docenko2011} of RbCs in a wide $R$-range. The tabulated $U^{ab}_i(R)$, $U^{db}_i(R)$, $d_{ij}^{ab}(R)$ and $L_{ij}^{ab}(R)$ functions could be found in the ASCII format in the Supplemented material~\cite{EPAPS}.

The details of the computational procedure can be found elsewhere~\cite{Pazyuk-2015}. Briefly, the inner core shell of both rubidium and cesium atoms was replaced by the shape-consistent non-empirical effective core potentials~\cite{Ross90} (ECP), leaving nine outer shells (eight sub-valence plus one valence) electrons for explicit treatment. The optimized molecular orbitals were constructed from the solutions of the state-averaged complete active space self-consistent field (SA-CASSCF) problem for all 18 electrons on the lowest (1-10)$^1\Sigma^+$ and (1-5)$^1\Pi$ electronic states taken with equal weights~\cite{Werner85}. The dynamical correlation was introduced by internally contracted multi-reference configuration interaction (MR-CI) method~\cite{Knowles92}, which was applied for only two valence electrons keeping the 16 sub-valence electrons frozen. The core-polarization potentials~\cite{CPP} (CPPs) of both atoms were implemented in order to take into account for the pronounced core-valence correlation effects. The CPP cut-off radii (see Table~\ref{CPP}) were adjusted to reproduce experimental non-relativistic energies~\cite{Atom} of the lowest excited Rb(5$^2$P) and Cs(5$^2$D) states, respectively.

\begin{table}[h]
\small
\caption{The static polarizability $\alpha_c$ [\onlinecite{AtomPol}] and cut-off radii $r_c$ used for the core-polarization potentials of the Rb and Cs atoms. All parameters in a.u.}\label{CPP}
\begin{tabular}{ccc}
\hline\hline
& $\alpha_c$ & $r_c$\\
\hline
Rb & 9.096  & 0.389\\
Cs & 15.687 & 0.243\\
\hline
\end{tabular}
\end{table}

\section{Results and Discussion}\label{discussion}
\subsection{Experimental term values and potential energy curves of the (3)$^1\Pi$ and (5)$^1\Sigma^+$ states}\label{PECs}

\begin{figure}
\includegraphics[scale=0.92]{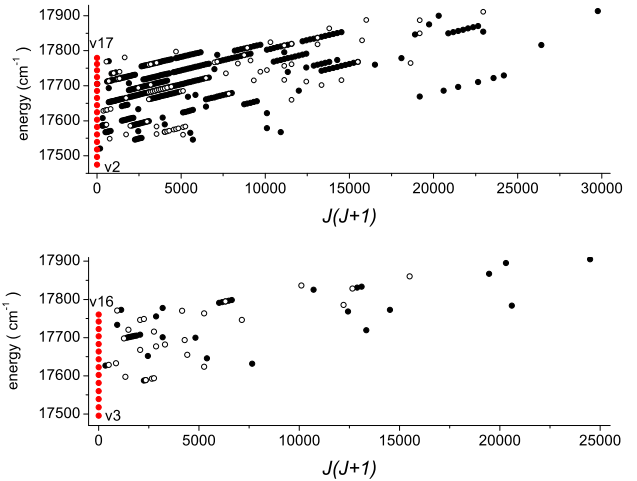}
\caption{Experimental term values data field available for the (3)$^1\Pi$ state: upper layer - $^{85}$Rb$^{133}$Cs, lower layer - $^{87}$Rb$^{133}$Cs. Empty circles denote $e$-symmetry levels, full circles denote $f$-symmetry levels. Red dots denote values extrapolated in Ref.~\cite{Kim94} to the lowest $J^{\prime}=1$ levels.}\label{Fig4_3Pi_data}
\end{figure}

\begin{figure}
\includegraphics[scale=0.95]{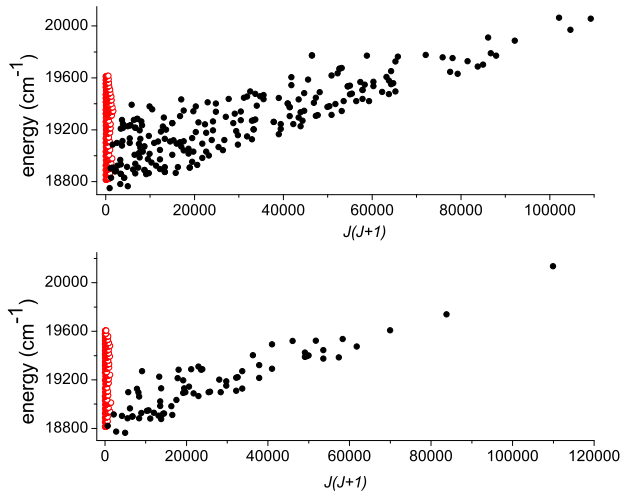}
\caption{Experimental term values data field available for the (5)$^1\Sigma^+$ state: upper layer - $^{85}$Rb$^{133}$Cs, lower layer - $^{87}$Rb$^{133}$Cs. Full circles denote the present experiment, red circles denote the low $J^{\prime}$-value levels observed in Ref.~\cite{Yoon-JCP2001}.}\label{Fig5_5Sig_data}
\end{figure}

Overall 493 rovibronic term values of the (3)$^1\Pi$ state and 292 term values of the (5)$^1\Sigma^+$ state were obtained in the present experiment. We expect that an uncertainty of measured term values is about 0.01 - 0.015 cm$^{-1}$; dominant contributions are the Doppler broadening of excitation transitions and the uncertainty of the calculated term values of the A$\sim$b complex. Figures~\ref{Fig4_3Pi_data} and \ref{Fig5_5Sig_data} show all experimental data set available for (3)$^1\Pi$ and (5)$^1\Sigma^+$ states of both $^{85}$Rb$^{133}$Cs and $^{87}$Rb$^{133}$Cs isotopologues. Besides the present data set for the (3)$^1\Pi$ state, we incorporated in the DPF analysis 29 rotationless ($J^{\prime}=1$) term values from Ref.~\cite{Kim94} for $v^{\prime}\in[2,17]$ and $v^{\prime}\in[3,17]$ vibrational levels of $^{85}$Rb$^{133}$Cs and $^{87}$Rb$^{133}$Cs, respectively. For the (5)$^1\Sigma^+$ state the originally measured absorption (5)$^1\Sigma^+ \leftarrow $ X $^1\Sigma^+$ line positions given in the EPAPS of Ref.~\cite{Yoon-JCP2001} were added to the corresponding energy of the ground state calculated with the highly accurate empirical $U^{emp}_X(R)$ PEC~\cite{Docenko2011}. Thus obtained 1196 rovibronic term values of both isotoplogues with $v^{\prime}\in[6,31]$ and $J^{\prime}\in[1,41]$ were involved into the DPF analysis along with the present (5)$^1\Sigma^+$ data.  The data from Ref.~\cite{Kim94} for the (3)$^1\Pi$ state and from Ref.~\cite{Yoon-JCP2001} for (5)$^1\Sigma^+$ state are represented in Figs.~\ref{Fig4_3Pi_data} and \ref{Fig5_5Sig_data} ,respectively, as vertical columns of points (in red).

\begin{figure}
\includegraphics[scale=0.9]{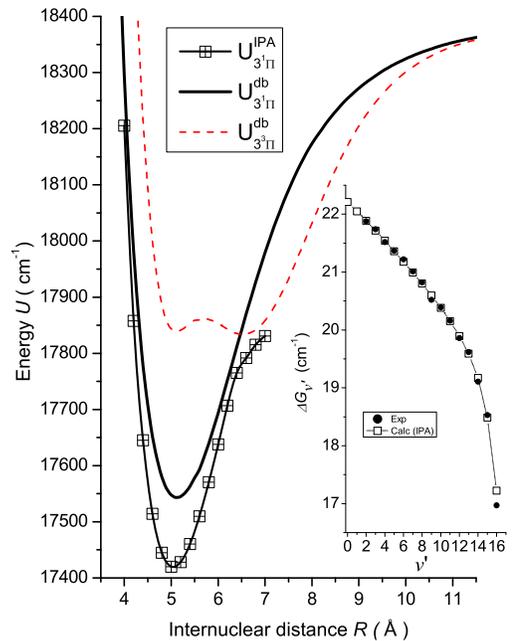}
\caption{Present empirical $U^{IPA}$ and difference-based $U^{db}$ adiabatic PECs constructed for (3)$^1\Pi$ and (3)$^3\Pi$ states. The inset displays the calculated and experimental~\cite{Kim94}
$\Delta G_{v^{\prime}} = E_{v^{\prime}+1} - E_{v^{\prime}}$ values as a function of the vibrational quantum number $v^{\prime}$.}\label{Fig6_IPA-1Pi}
\end{figure}

\begin{figure}
\includegraphics[scale=0.95]{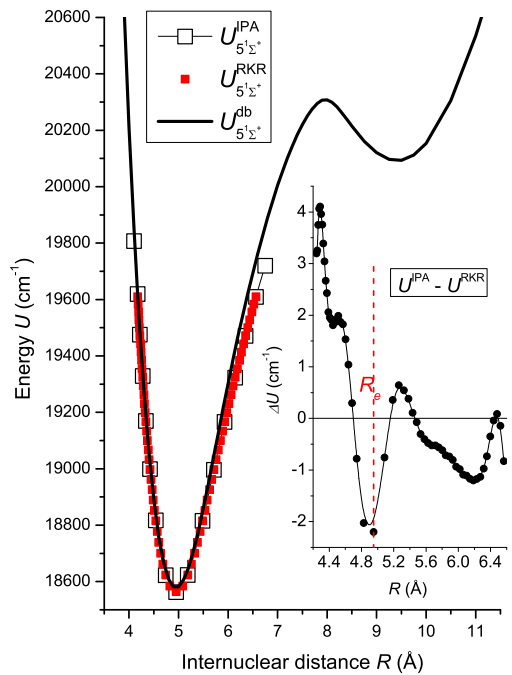}
\caption{Present empirical $U^{IPA}$ and difference-based $U^{db}$ adiabatic PECs constructed for the (5)$^1\Sigma^+$ state. The RKR potential from Ref.~\cite{Yoon-JCP2001} is also presented. The inset represents the difference between the present IPA and RKR potentials.}\label{Fig7_IPA-1Sig}
\end{figure}

The resulting IPA PECs of both states are listed in Table~\ref{PECIPA} and depicted, along with the "difference-based" PECs, in Fig.~\ref{Fig6_IPA-1Pi} and  Fig.~\ref{Fig7_IPA-1Sig}, respectively. The equilibrium parameters $R_e$ and $T_e$ of the present PECs are compared with the available literature data in Table~\ref{PECcompare}. One may note that the present $R_e$ and $T_e$ values are close enough to the preceding \emph{ab initio} and empirical data for both states.

\begin{figure}
\includegraphics[scale=0.92]{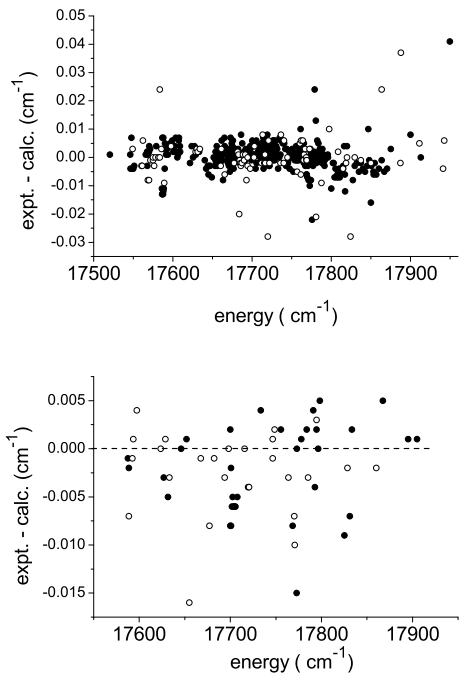}
\caption{Residuals $E^{expt}$ - $E^{IPA}$ for (3)$^1\Pi$ state rovibronic levels observed in the present experiment. Empty circles denote $e$-levels, full circles denote $f$-levels. Upper layer - $^{85}$Rb$^{133}$Cs, lower layer -  $^{87}$Rb$^{133}$Cs.}\label{Fig7}
\end{figure}

\begin{figure}
\includegraphics[scale=0.86]{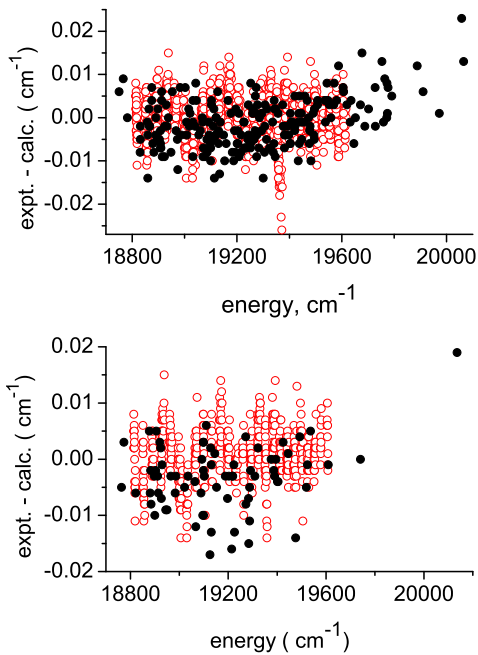}
\caption{Residuals $E^{expt}$ - $E^{IPA}$ for the (5)$^1\Sigma^+$ state rovibronic levels included in the fit: black full circles - present experiment, red empty circles - data from Ref.~\cite{Yoon-JCP2001}. The upper layer -  $^{85}$Rb$^{133}$Cs, lower layer -  $^{87}$Rb$^{133}$Cs.}\label{Fig8}
\end{figure}

The present IPA potentials reproduce the observed term values for the (3)$^1\Pi$ and (5)$^1\Sigma^+$ states with experimental spectroscopic accuracy as can be seen in Figs.~\ref{Fig7} and ~\ref{Fig8}, respectively. More than 95\%
of the present FTS data obtained for the (3)$^1\Pi$ state are reproduced within $\pm$ 0.01 cm$^{-1}$. It is interesting that for $^{87}$Rb$^{133}$Cs (lower layer in Fig.~\ref{Fig7}) some slight systematic shift about -0.003 cm$^{-1}$ can be observed. The residuals for the (3)$^1\Pi$ state term values from Ref.~\cite{Kim94} are presented in EPAPS~\cite{EPAPS}; they are substantially larger in accordance with reported in with reported Ref.~\cite{Kim94} experimental inaccuracy of 0.1 cm$^{-1}$. The vibrational quanta of the (3)$^1\Pi$ state $\Delta G_{v^{\prime}} = E_{v^{\prime}+1} - E_{v^{\prime}}$ abruptly decrease for $v^{\prime}\ge 14$ levels (see the inset in Fig.~\ref{Fig6_IPA-1Pi}) yielding a so-called  "shelf" in the corresponding adiabatic PEC. Such a behaviour can be attributed to avoided-crossing effect caused by the local spin-orbit coupling with the nearby triplet (3)$^3\Pi$ state.

The residuals for the (5)$^1\Sigma^+$ state also demonstrate high accuracy of the present IPA PEC which is, as expected,  close to the preceding RKR potential from Ref.~\cite{Yoon-JCP2001} (see the inset of Fig.~\ref{Fig7_IPA-1Sig}). It should be mentioned however, that the RKR PEC ~\cite{Yoon-JCP2001} reproduces the experimental term values with substantially less accuracy, especially outside of the observed range of small $J^{\prime}$-values; the corresponding figure is presented in the EPAPS~\cite{EPAPS}.
It should be noted that the present FTS measurements significantly extend the range of experimental term values corresponding to high and intermediate rotational quantum numbers $J^{\prime}$ of both studied states. The range of observed vibrational levels was also expanded down to $v^{\prime}= 0$ for the (3)$^1\Pi$ state and to $v^{\prime}= 1$ for the (5)$^1\Sigma^+$ state.

\begin{table}[t!]
\caption{Mass-invariant IPA potentials obtained for the (5)$^1\Sigma^+$ and (3)$^1\Pi$ states of the RbCs molecule. For the doubly degenerated (3)$^1\Pi$ state the IPA points correspond only to the "unperturbed" $f$-component. }\label{PECIPA}
\begin{center}
\begin{tabular}{cccc}
\hline
\hline
   R   & $U^{IPA}_{(5)^1\Sigma^+}$ & R & $U^{IPA}_{(3)^1\Pi}$\\
   \AA & cm$^{-1}$ & \AA & cm$^{-1}$\\
\hline
4.1000 &	19807.379  &  4.0	  & 18205.267\\
4.1732 &	19619.034  &  4.2	  & 17857.830\\
4.2206 &	19475.817  &  4.4	  & 17645.290\\
4.2751 &	19328.445  &  4.6	  & 17514.442\\
4.3414 &	19169.777  &  4.8	  & 17444.714\\
4.4250 &	18999.012  &  5.0	  & 17419.908\\
4.5388 &	18817.032  &  5.2	  & 17428.181\\
4.7395 &	18622.485  &  5.4	  & 17460.458\\
4.9510 &	18562.362  &  5.6	  & 17509.820\\
5.1866 &	18623.620  &  5.8	  & 17570.475\\
5.4836 &	18815.085  &  6.0	  & 17637.708\\
5.7135 &	18996.575  &  6.2	  & 17706.632\\
5.9310 &	19165.930  &  6.4	  & 17764.997\\
6.1447 &	19323.213  &  6.6	  & 17791.839\\
6.3553 &	19470.917  &  6.8	  & 17815.424\\
6.5625 &	19609.251  &  7.0	  & 17831.022\\
6.7500 &	19720.086\\
\hline
\hline
\end{tabular}
\end{center}
\end{table}

\begin{table}[t!]
\caption{Comparison of equilibrium distance $R_e$ (in \AA) and electronic energy $T_e$ (in cm$^{-1}$) available for the (5)$^1\Sigma^+$ and (3)$^1\Pi$ states of RbCs molecule. The theoretical results correspond to
pure Hund's (a) coupling case. The abbreviation PW marks the present work.}\label{PECcompare}
\begin{center}
\begin{tabular}{lcccc}
\hline
\hline
  & \multicolumn{2}{c}{(5)$^1\Sigma^+$} & \multicolumn{2}{c}{(3)$^1\Pi$}\\
\hline
  Source   & $R_e$ & $T_e$ & $R_e$ & $T_e$\\
\hline
  Theor. [\onlinecite{Allouche00}] & 4.75 & 18902 & 5.04 & 17065\\
  Theor. [\onlinecite{Pavolini89}] & 4.87 & 18562 & 5.06 & 17633\\
  Theor. [\onlinecite{Lim06}] & 4.97 & 18481 & 5.24 & 17542\\
  Theor. [\onlinecite{Zaitsevskii05}] & 4.95 & 18551 & 5.07 & 17598\\
  Theor. [PW]                 & 4.95 & 18578 & 5.10 & 17542\\
  \hline
  Expt. [\onlinecite{Yoon-JCP2001, Kim94}]& 4.951 & 18564.6  &  & 17418.9\\
  Expt.  [PW]          & 4.945 & 18562.33 & 5.041 & 17419.17\\
\hline
\hline
\end{tabular}
\end{center}
\end{table}

\subsection{The $q$-factors of the (3)$^1\Pi$ state}\label{q-factor}

\begin{figure}
\includegraphics[scale=0.9]{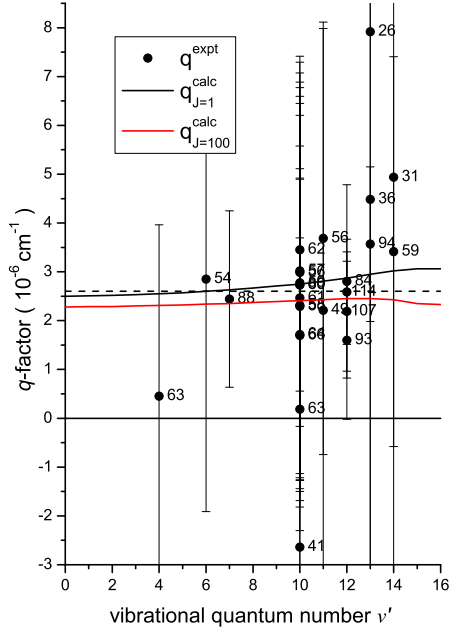}
\caption{Experimental and calculated $q$-factors in the (3)$^1\Pi$ state of the $^{85}$Rb$^{133}$Cs molecule. Rotational quantum number $J^\prime$ of measured levels is presented in the figure. Dashed horizontal line corresponds to the averaged experimental $q$-value 2.6$\times$10$^{-6}$ cm$^{-1}$.}\label{Fig3_q-factor}
\end{figure}

For several levels of the double degenerated (3)$^1\Pi$ state the experimental term values of both $e$ and $f$ components were determined for the same $v^{\prime}$, $J^{\prime}$ levels and the corresponding $\Delta^{expt}_{e/f}(v^{\prime},J^{\prime})=E^e_{^1\Pi} - E^f_{^1\Pi}$ splitting was obtained, see Fig.~\ref{Fig3_q-factor}. In accordance with the second order perturbation theory, the observed ${e/f}$-splitting can be approximated as
\begin{eqnarray}\label{efsplitting}
\Delta^{expt}_{e/f}=q^{expt}[J^{\prime}(J^{\prime}+1)-1],
\end{eqnarray}
where the experimental $q^{expt}$-factor is weakly dependent on $v^{\prime}$ and $J^{\prime}$ values. Because of great statistical errors of $q^{expt}$-values $v^{\prime},J^{\prime}$-dependence can not be determined correctly from the present experiment. Averaging of all $q^{expt}$ data yields value about $2.6\times10^{-6}$ cm$^{-1}$.

In the case of a \emph{regular} perturbation~\cite{Fieldbook-2004} the \emph{ab initio} evaluated $Q(R)$ function (\ref{Qfunction}) and the empirical parameter $s=1.0455$ obtained during the present DPF analysis should be  unambiguously related to the experimental $q$-factors due to the approximate sum rules~\cite{PupyshevCPL94}
\begin{eqnarray}\label{qfactor}
q^{calc}&\approx &2\sum_{^1\Sigma^+}\sum_{v_{^1\Sigma^+}}\frac{|\langle v^{J^{\prime}}_{^1\Pi}|BL_{^1\Pi-^1\Sigma^+}|v^{J^{\prime}}_{^1\Sigma^+}\rangle|^2}{E^{J^{\prime}}_{v_{^1\Pi}}-E^{J^{\prime}}_{v_{^1\Sigma^+}}}\nonumber\\
&\approx &s\langle v^{J^{\prime}}_{^1\Pi}|BQB|v^{J^{\prime}}_{^1\Pi}\rangle .
\end{eqnarray}
Indeed, Figure~\ref{Fig3_q-factor} demonstrates overall good agreement between the present measured $q^{expt}$ and calculated $q^{calc}$ values.

The value of the (3)$^1\Pi$ $q$-factor ($\approx 2.6\times10^{-6}$ cm$^{-1}$) appeared to be about 3-4 times larger than those estimated for the lower~\cite{Gustavsson88CPL} (2)$^1\Pi$ and higher~\cite{Zaitsevskii05} (4)$^1\Pi$ lying states, namely: $q_{(2)^1\Pi}\approx 5.9\times10^{-7}$ and $q_{(4)^1\Pi}\approx 8.0\times10^{-7}$ cm$^{-1}$. This difference could be attributed to the dominant contribution of the single (4)$^1\Sigma^+$ state into the sum (\ref{Qfunction}). Furthermore, both relations (\ref{efsplitting}) and (\ref{qfactor}) are valid only for low-lying vibrational levels of the (3)$^1\Pi$ state since the $Q(R)$ function has a singularity near the point $R\approx 4.2$~\AA, where the repulsive walls of adiabatic (3)$^1\Pi$ and (4)$^1\Sigma^+$ PECs are crossing each other (see Fig.~\ref{Fig1_PECs}).

\subsection{Radiative lifetimes and branching ratios of the (3)$^1\Pi$ and (5)$^1\Sigma^+$ states}\label{lifetime}

The present \emph{ab initio} adiabatic PECs $U^{ab}_i(R)$ and transition dipole moments $d_{ij}^{ab}(R)$, see  Section~\ref{abinitio}, were used to estimate a radiative lifetime $\tau_i$ of the upper (3)$^1\Pi$ and (5)$^1\Sigma^+$ states along with their vibronic branching ratios $R_{if}$ into the lower-lying singlet state manifold. To avoid tedious summation over bound part and integration over continuum part of the vibrational spectra of the lower states we have used the approximate sum rule~\cite{PupyshevCPL94}.
\begin{eqnarray}\label{tausum}
\frac{1}{\tau_i}\approx\frac{8{\pi }^2}{3\hbar {\epsilon_0}}
\langle v_i^{J^{\prime}}|\sum_j [\Delta U^{ab}_{ij}]^3[d_{ij}^{ab}]^2|v_i^{J^{\prime}}\rangle,
\end{eqnarray}
\begin{eqnarray}\label{brsum}
R_{if} = \frac{\langle v_i^{J^{\prime}}|[\Delta U^{ab}_{if}]^3[d^{ab}_{if}]^2|v_i^{J^{\prime}}\rangle}{\langle
v_i^{J^{\prime}}|\sum_j [\Delta U^{ab}_{ij}]^3[d^{ab}_{ij}]^2|v_i^{J^{\prime}}\rangle},
\end{eqnarray}
where $\Delta U^{ab}_{ij}(R)=U^{ab}_i(R)-U^{ab}_j(R)$ is the difference of the \emph{ab initio} PECs, while the corresponding vibrational wavefunctions $|v^{J^{\prime}}_i\rangle$ of the upper states were calculated using the present IPA potentials.

\begin{figure}
\includegraphics[scale=0.23]{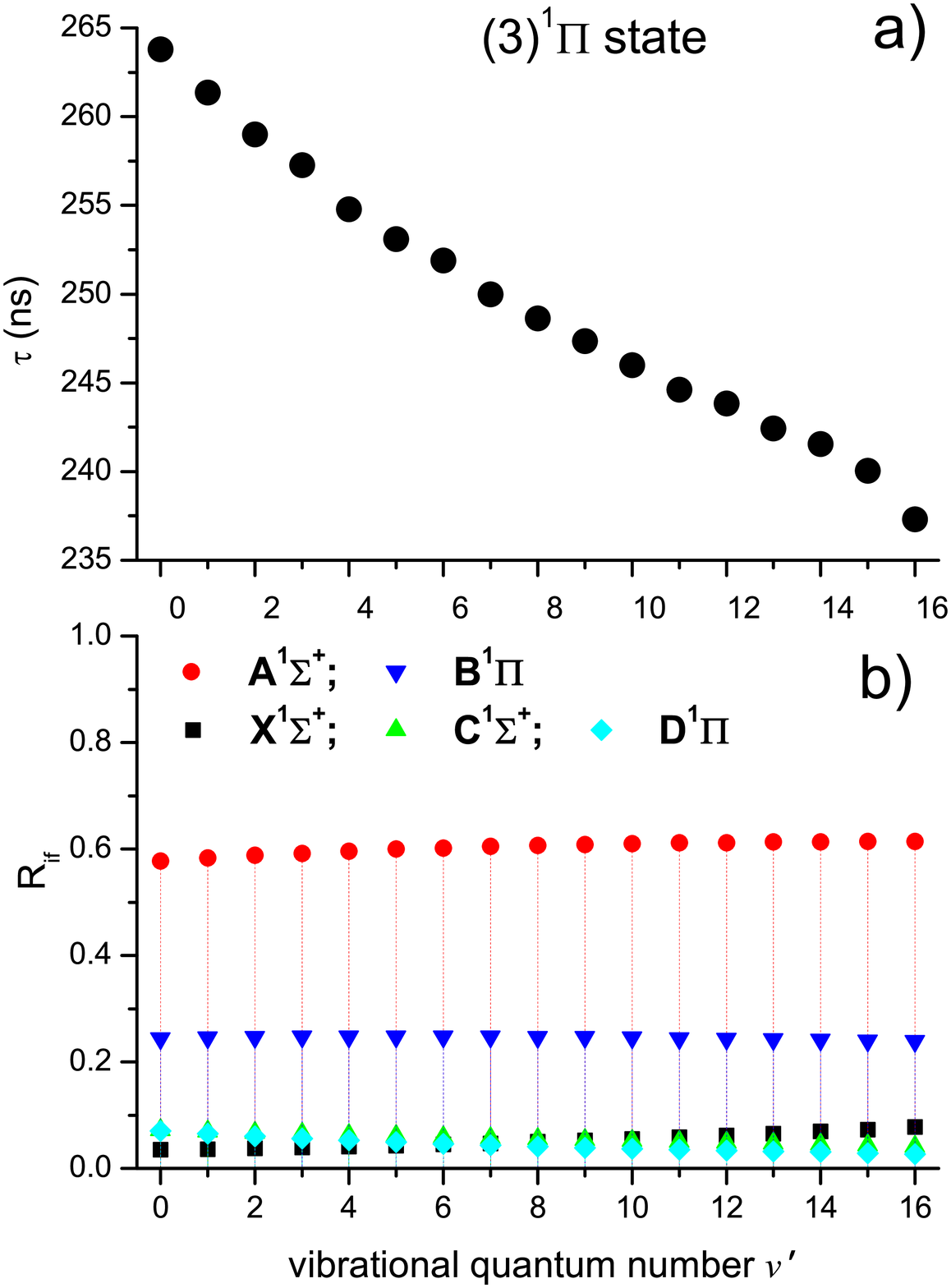}
\caption{Radiative lifetimes (\textbf{a}) and electronic branching ratios (\textbf{b}) of the $^{85}$Rb$^{133}$Cs (3)$^1\Pi$ state as dependent on the vibrational quantum number
evaluated according to the relations (\ref{tausum}) and (\ref{brsum}), respectively.}\label{Fig13}
\end{figure}

\begin{figure}
\includegraphics[scale=0.26]{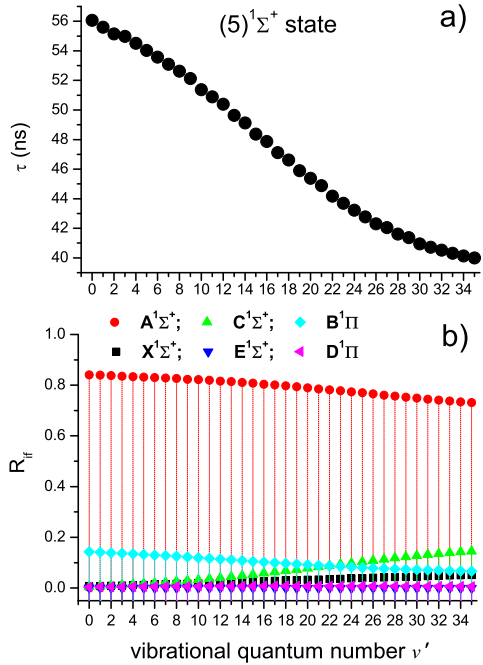}
\caption{Radiative lifetimes $\tau$ (\textbf{a}) and electronic branching ratios $R_{if}$ (\textbf{b}) on the $^{85}$Rb$^{133}$Cs (5)$^1\Sigma^+$ state as dependent of the vibrational quantum number
evaluated according to the relations (\ref{tausum}) and (\ref{brsum}), respectively.}\label{Fig14}
\end{figure}

The resulting  $\tau_i$ and $R_{if}$ values for the (3)$^1\Pi$ and (5)$^1\Sigma^+$ states are depicted in Fig.~\ref{Fig13} and Fig.~\ref{Fig14}, respectively. The calculated radiative lifetimes of both states rapidly decrease as the vibrational quantum number $v^{\prime}$ increases. The $\tau$-values predicted for the vibrational levels of the (5)$^1\Sigma^+$ state are about 5-6 times smaller than those for the rather long living (3)$^1\Pi$ state.

According to the calculations the dominant decay channel for both $i\in [(5)^1\Sigma^+;(3)^1\Pi]$ states is the optical transition to the lower A$^1\Sigma^+$ state (see Fig.~\ref{Fig13} and Fig.~\ref{Fig14}) due to a large value of the relevant spin-allowed dipole moments (see EPAPS~\cite{EPAPS}). The contribution of the (5)$^1\Sigma^+$;(3)$^1\Pi\to$B$^1\Pi$ transitions into the total decay does not exceed 30\% while all other states (including the lowest ground state) contribute almost nothing. These estimates explain the fact that during the present LIF experiments we were not able to observe, except a few cases, very week (3)$^1\Pi\to$X$^1\Sigma^+$ and (5)$^1\Sigma^+\to$X$^1\Sigma^+$ transitions hidden in the bush of strong (4)$^1\Sigma^+\to$X$^1\Sigma^+$ progressions.

\subsection{Intensity distributions in the (3)$^1\Pi\to$A$\sim$b and (5)$^1\Sigma^+\to$A$\sim$b LIF progressions}\label{intensity}

\begin{figure}
\includegraphics[scale=0.9]{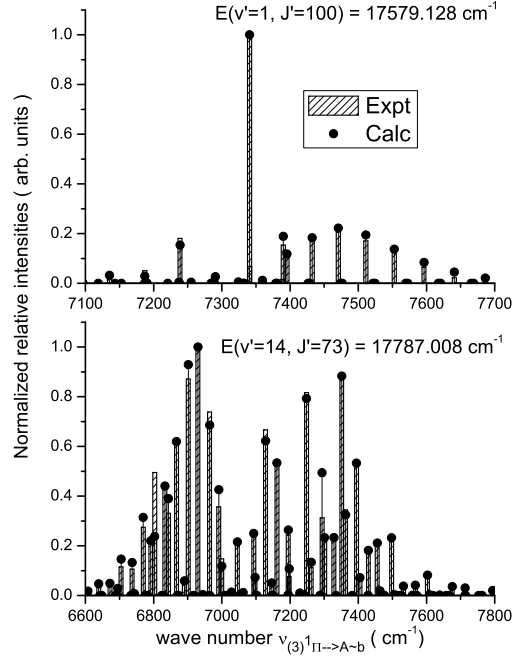}
\caption{Experimental $I^{expt}$ and calculated $I^{calc}$ relative intensity distributions in the singlet ($J^{\prime}=J^{\prime\prime}$) LIF progressions with transitions from the particular $v^{\prime}=1$ (upper layer) and $v^{\prime}=14$ (lower layer) vibrational levels of the (3)$^1\Pi$ state to the lower-lying levels of the A$\sim$b complex of the $^{85}$Rb$^{133}$Cs isotopologue.}\label{Fig11}
\end{figure}

\begin{figure}
\includegraphics[scale=0.9]{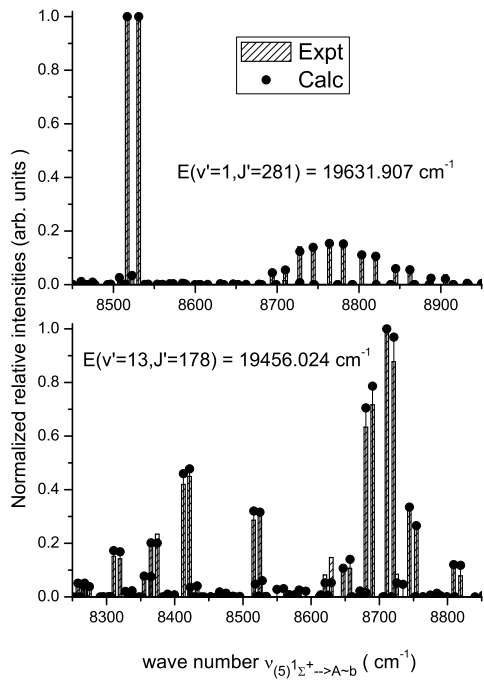}
\caption{The experimental $I^{expt}$ and calculated $I^{calc}$ relative intensity distribution in the doublet ($J^{\prime\prime}=J^{\prime}\pm 1$) LIF progressions from $v^{\prime}=1$ (upper layer) and $v^{\prime}=13$ (lower layer) vibrational levels of the (5)$^1\Sigma^+$ state to rovibronic levels of the A$\sim$b complex of the $^{85}$Rb$^{133}$Cs isotopologue.}\label{Fig12}
\end{figure}

The accuracy of the derived IPA potentials of both (5)$^1\Sigma^+$ and (3)$^1\Pi$ states as well as \emph{ab initio} (5)$^1\Sigma^+$-A$^1\Sigma^+$ and (3)$^1\Pi$-A$^1\Sigma^+$ transition dipole moments was additionally checked  by a comparison  of the simulated and measured relative intensity distributions in the particular (3)$^1\Pi\to$A$\sim$b and (5)$^1\Sigma^+\to$A$\sim$b LIF progressions. The relevant rovibronic transition probabilities from the upper $i$-state ($i\in (3)^1\Pi; (5)^1\Sigma^+$) to a lower-lying levels of the perturbed A$\sim$b complex were calculated according to the relation:
\begin{eqnarray}\label{Iteninterfer}
I_{i\to A\sim b}^{Calc} &\sim& \nu^4_{i\to A\sim b}|\langle v^{J^{\prime}}_i|d^{ab}_{i-A}|\phi_A^{J^{\prime\prime}}\rangle|^2,\\
\nu_{i\to A\sim b} &=& E_{v_i^{\prime}}^{J^{\prime}}-E^{J^{\prime\prime}}_{A\sim b}, \nonumber
\end{eqnarray}
where $\phi_A^{J^{\prime\prime}}(R)$ are the non-adiabatic vibrational wavefunctions belonging to the singlet component of the A$\sim$b complex while $E_{v_i^{\prime}}^{J^{\prime}}$, $|v^{J^{\prime}}_i\rangle$ are the adiabatic vibrational eigenvalues and eigenfunctions of the upper state calculated with the present IPA potentials.  The comparison of the theoretical intensity distributions $I^{calc}$ with their experimental counterparts $I^{expt}$ is presented in Fig.~\ref{Fig11} and Fig.~\ref{Fig12}. The observed and calculated intensities agree well enough, thus demonstrating the reliability of the performed analysis.

\section{Conclusions}\label{conclusion}
We have demonstrated that the well-established perturbed levels of the singlet-triplet A$^1\Sigma^+\sim$~b$^3\Pi$ complex of the RbCs molecule can be successfully used for unambiguous rotational assignment of the high resolution FTS LIF progressions originating from the highly excited electronic states. The experimental rovibronic term values of the (3)$^1\Pi$ and (5)$^1\Sigma^+$ states were determined in a wide range of the rotational quantum number $J^{\prime}$. The range of observed vibrational levels was expanded down to $v^{\prime}= 0$ for (3)$^1\Pi$ state and $v^{\prime}= 1$ for (5)$^1\Sigma^+$ state. The obtained data were involved, together with the preceding term values~\cite{Kim94, Yoon-JCP2001}, in a direct-potential-fit analysis performed in the framework of inverted perturbation approach. The simulated intensity distributions in the observed LIF progressions are found to be remarkably close to their experimental counterparts. A good agreement between the experimental and calculated $q-$factors validates the \emph{unique perturber} approximation~\cite{Fieldbook-2004} for the (3)$^1\Pi$ state. Based on the \emph{ab initio} transition dipole moments estimate of lifetimes and vibronic branching ratios confirms a dominant contribution of (5)$^1\Sigma^+ \rightarrow$A$^1\Sigma^+$ and (3)$^1\Pi\rightarrow$A$^1\Sigma^+$ transitions into total radiative decay of the both upper states.

\begin{acknowledgments}
Moscow team thanks for the support by the RFBR grant No.16-03-00529a. Riga team acknowledges support from University of Latvia funding:  Base/Performance No A5-AZ27, Y9-B013, and support from the EU project Laserlab - Europe H2020 EC - GA654148.
\end{acknowledgments}.

\end{document}